\documentclass[12pt,preprint]{emulateapj}

\newcommand{\apg}{^{>}_{\sim}}
\newcommand{\apll}{^{<}_{\sim}}

\shorttitle{The Gas Reservoir around a Star Forming Galaxy}
\shortauthors{Frye, et al.}

\begin{document}



\title{Observations of the Gas Reservoir around a Star Forming 
Galaxy in the Early Universe}


\author{Brenda L. Frye\altaffilmark{1}, David V. Bowen\altaffilmark{2}, 
Mairead Hurley\altaffilmark{1}, Todd M. Tripp\altaffilmark{3}, Xiaohui Fan\altaffilmark{4},
Bradley Holden\altaffilmark{5}, Puragra Guhathakurta\altaffilmark{5}, 
Dan Coe\altaffilmark{6}, Tom Broadhurst\altaffilmark{7}, Eiichi Egami\altaffilmark{4},
G. Meylan\altaffilmark{8}}

\altaffiltext{1}{School of Physical Sciences, Dublin City University, Glasnevin, Dublin 9, Ireland}
 \altaffiltext{2}{ Department of Astrophysical Sciences, Peyton Hall, Princeton University, Princeton, NJ  08540}
 \altaffiltext{3}{ Department of Astronomy, University of Massachusetts, Amherst, MA  01003}
 \altaffiltext{4}{ Department of Astronomy and Steward Observatory, University of Arizona, Tucson, AZ  85721} 
\altaffiltext{5}{UCO/Lick Observatory, University of California, Santa Cruz, CA  95064}
\altaffiltext{6}{Jet Propulsion Laboratory, M/S 169-327, 4800 Oak Grove Drive, Pasadena, CA  
91109}
\altaffiltext{7}{School of Physics and Astronomy, Tel Aviv University, Ramat Aviv 69988, Israel}
\altaffiltext{8}{Laboratoire d'Astrophysique, Ecole Polytechnique F\'ed\'erale de Lausanne (EPFL) Observatoire, CH-1290, Sauverny, Suisse}



\begin{abstract}
We present a high signal-to-noise spectrum of a bright galaxy at $z$ = 4.9
in 14 h of integration on VLT FORS2.  This galaxy is extremely bright,
$i_{850} = 23.10 \pm 0.01$, and is strongly-lensed by the foreground 
massive galaxy cluster Abell 1689 ($z=0.18$).  Stellar continuum is
seen longward of the Ly$\alpha$ emission line at  $\sim7100$ \AA, 
while intergalactic H~I produces strong absorption shortward of Ly$\alpha$.
Two transmission spikes at $\sim$6800 \AA \ and $\sim$7040 \AA \ are 
also visible, along with other structures at shorter wavelengths.   
Although fainter than a QSO, the absence of a strong central 
ultraviolet flux source in this star forming galaxy enables a measurement 
of the H~I flux transmission in the intergalactic medium (IGM) in the 
vicinity of a high redshift object.  We find that the effective H I optical 
depth of the IGM is remarkably high within a large 14 Mpc (physical) 
region surrounding the galaxy  compared to that seen towards QSOs 
at similar redshifts.  Evidently, this high-redshift galaxy is located in a 
region of space where the amount of H~I is much larger than that seen
at similar epochs in the diffuse IGM.  We argue that observations of 
high-redshift galaxies like this one provide unique insights on the 
nascent stages of baryonic large-scale structures that evolve into 
the filamentary cosmic web of galaxies and clusters of galaxies 
observed in the present universe.
\end{abstract}






\section{Introduction}
Hydrodynamic simulations tell us that dark matter near the epoch
of galaxy formation 
collapses into an ordered filamentary pattern, the so-called ``cosmic web."
In turn, this cosmic web is thought to cradle high-redshift galaxies in dense 
nodes that are opaque to the extragalactic ultraviolet background radiation.
Observations of
H~I surrounding these high-redshift objects provide information on the 
likely reservoirs from which galaxies assemble their gas.  Hitherto, it has only been
possible to measure H~I opacities towards bright Quasi-stellar
Objects (QSOs); unfortunately, the high ultraviolet flux from the
QSOs  ionizes the hydrogen clouds in their vicinity, thereby making  the 
determination of H~I cloud physical conditions unreliable close to the QSO.

New techniques and facilities in the past decade have enabled
detailed observational studies of the IGM in the
early universe ($z > 5$).  For example, the spectra of high-redshift 
QSOs show a plethora of H~I Ly$\alpha$ absorption lines, the ``Ly$\alpha$ forest",
as well as lines from a wide variety of heavier elements, all of
which provide detailed information about the high-redshift IGM along the
line-of-sight towards the QSO. Around
the QSOs themselves, however, the ultraviolet flux is so high that it typically ionizes
the surrounding gas.   Thus
QSOs are not ideal for studying the pervasive IGM close to the QSO.
Instead, we suggest that bright galaxies may provide 
better probes for the study of IGM conditions
in the proximity of high-redshift objects, as these objects emit fewer ultraviolet
photons than QSOs do.

The subject of this paper is the Ly$\alpha$ forest in the spectrum of an
unusually bright high-redshift galaxy at $z = 4.866$ (Frye et al. 2002, 2007).  
The galaxy is situated behind the massive galaxy
cluster Abell 1689 ($z = 0.183$), which magnifies the starlight of this
background object by a factor of 10.3 by the effect of
gravitational lensing (Broadhurst et al. 2005).  
Multicolor Hubble Space Telescope (HST) imaging of one of the faint lensed 
images, hereafter designated $A1689\_7.1$, is shown in Figure 1.  
We assume a cosmology for this paper of 
$H_0 = 70$ km s$^{-1}$ Mpc$^{-1}$, $\Omega_{m,0} = 0.3$, and 
$\Omega_{\Lambda,0} = 0.7$.

\begin{figure}
\includegraphics[viewport=60 -10 250 350,scale=0.7]{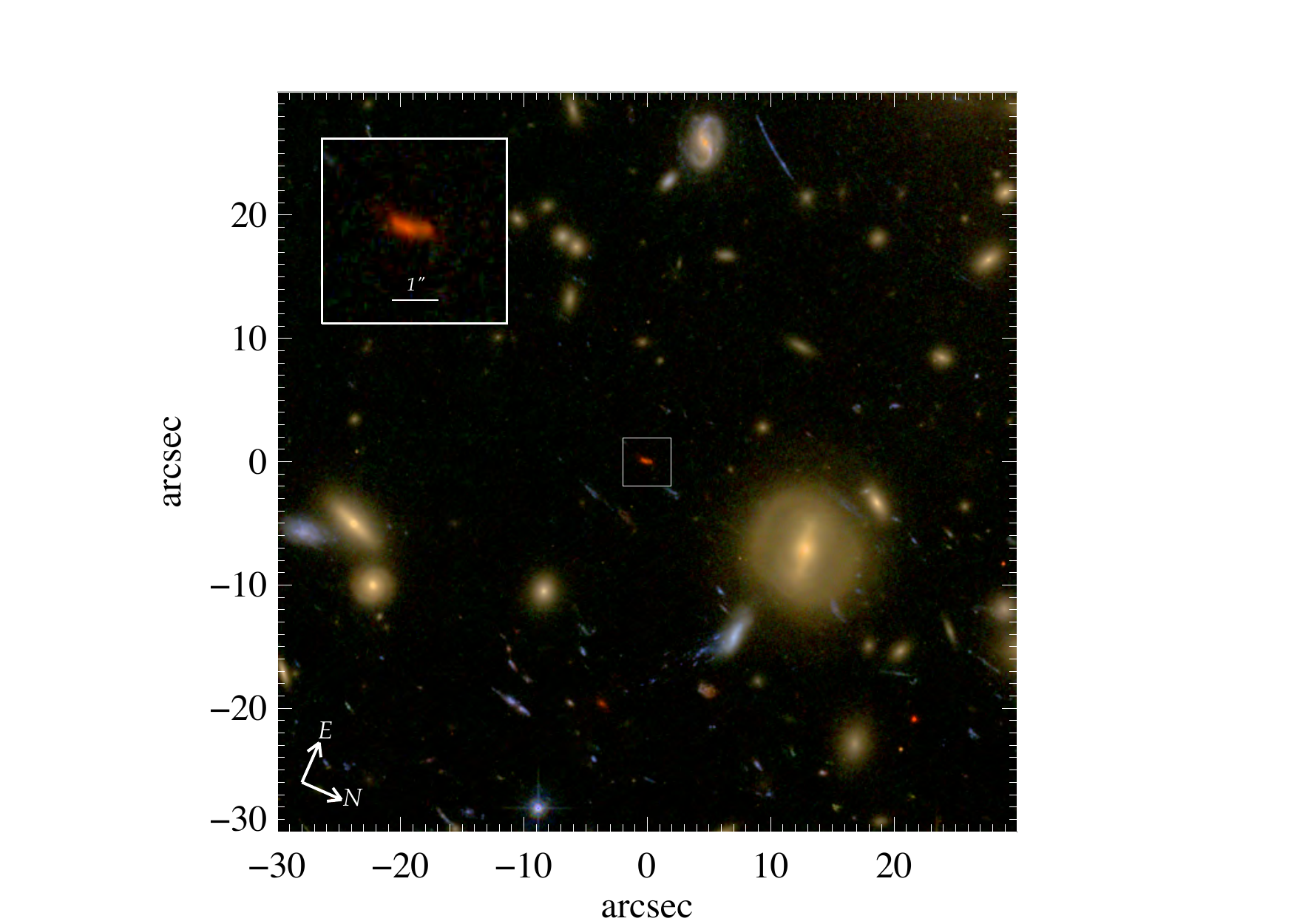}
\caption{HST Advanced Camera for Surveys (ACS) $gri$ optical photo of  $A1689\_7.1$ 
and surrounding region near the center of the galaxy cluster Abell 1689 ($z=0.18$).
A single galaxy at $z=4.9$ has been gravitationally-lensed by this foreground
galaxy cluster (with member galaxies visible here as yellow spheroid-shaped objects) 
into a system of three arclets; $A1689\_7.1$ is the brightest of the three and is 
shown above and in the inset.  This arclet is extremely bright, $i_{775}=23.10 \pm 0.01$ 
magnitudes, after being magnified by a factor of $10.3$.  $A1689\_7.1$  is also spatially 
resolved, with an estimated unlensed angular size of 0.094\arcsec,  corresponding to an 
intrinsic linear size of only 600 pc.  $A1689\_7.1$ was discovered as a part of a 
ground-based galaxy redshift survey (Frye et al. 2002, 2007). \label{fig_image}}
\end{figure}

\begin{figure}
\includegraphics[viewport = 0 0 700 520, scale=0.35]{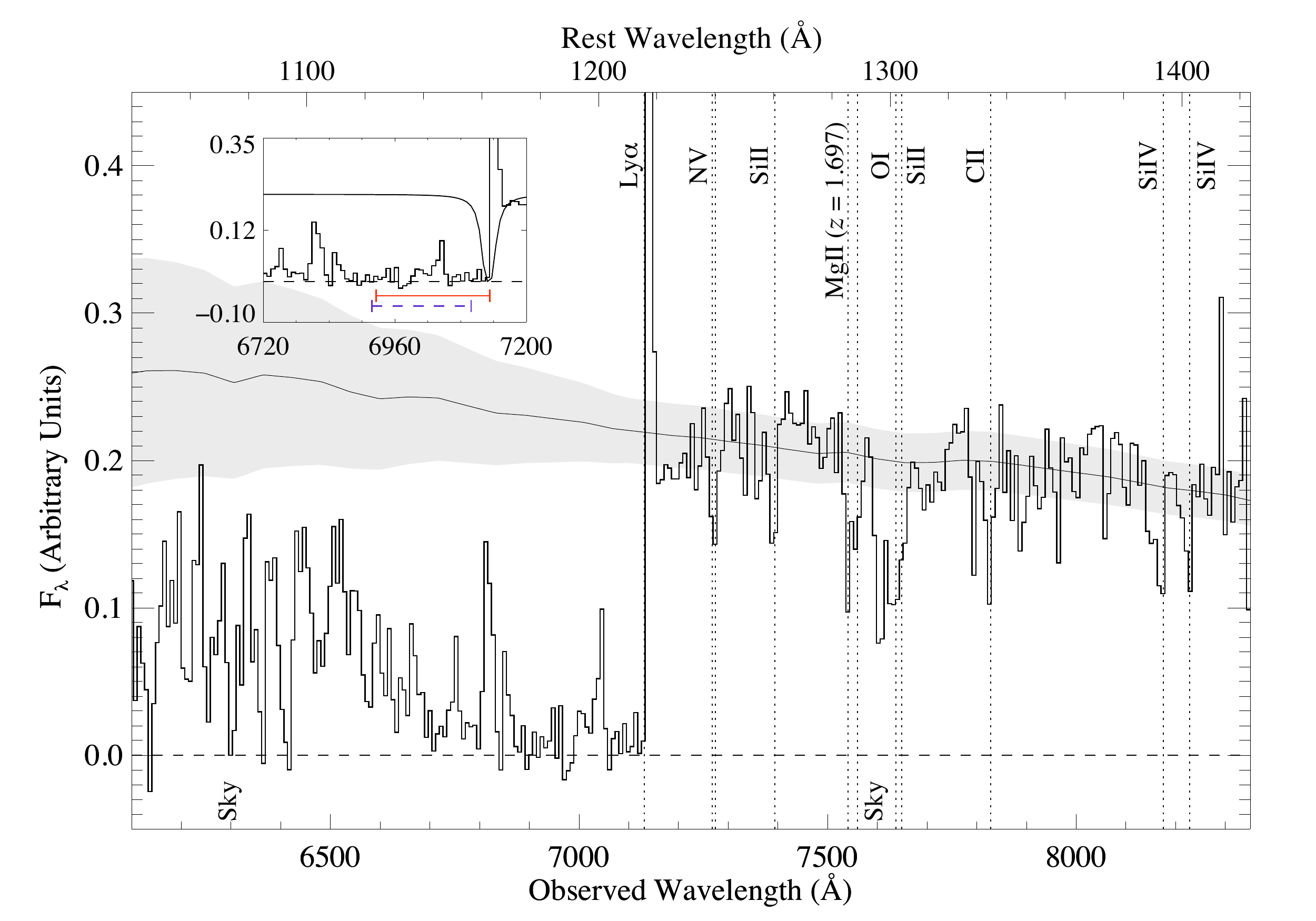}
\caption{VLT FORS2 spectrum of  $A1689\_7.1$ at $z=4.866$.  The stellar continuum 
of the galaxy is seen longward of the Ly$\alpha$ emission line at $\sim7100$ \AA, 
while intergalactic H~I produces strong absorption shortward of Ly$\alpha$.
Two transmission spikes at $\sim$6800 \AA \ and $\sim$7040 \AA \ are also visible, 
along with other structures at shorter wavelengths.  The data enable a measurement 
of the flux transmission along the line of sight to this star-forming galaxy. 
The best-fit \citet{Bruzual:03} model spectrum without corrections for dust extinction
and attenuation by the intervening Ly$\alpha$ forest is overlaid, including 
a gray region marking our 1$\sigma$ continuum-placement uncertainties.  
We select a Chabrier inital mass function (Padova 1994), stellar evolution tracks, 
and solar metallicity for the model, and give the details regarding the fitting procedure in \S4.
The inset figure shows a theoretical  H~I Ly$\alpha$ profile absorption fit
 to the data, with horizontal bars delimiting the physical extents of the proximity zone 
 for a QSO (red solid line), and the extent of the first of the five equally-spaced 
 redshift bins in our study over which we measure the H~I opacity (blue-dashed line), 
 respectively.  Owing to the lack of a significant proximity zone for our galaxy, 
a new physical scale is probed close to a high redshift object within a $\sim14$ 
physical Mpc radius (blue dashed line).\label{fig_spec }}
\end{figure}

\section{Observations and Spectrum}

Our spectrum of $A1689\_7.1$ is made possible with present telescopes
 only by the combination of strong lensing {\it and} 
unusually-long observations.  We obtained the spectrum for $A1689\_7.1$ in 14 hours at the 
Very Large Telescope (VLT) with the Focal Reducer and 
low-dispersion Spectrograph (FORS2) in 2001 June and July 
and report a spectral resolution of 
$R = \lambda/\Delta\lambda \approx 530$\footnote{The observations were made
on VLT/Kueyen FORS2 under observing programme 67.A-0618B}.  We fluxed the data
with the standard star LTT9239, and performed the reductions with a code written by the
first author in IDL designed to maximize the signal to noise of background-limited
objects.  See \citet{Frye:07} for more details on this purpose-built spectroscopic reduction package.


The spectrum of $A1689\_7.1$ (Figure~2) shows an  H~I Ly$\alpha$ emission line
and several strong low- and high-ionization metal absorption lines detected
against the stellar continuum.  
On the basis of a mean velocity difference
of $v$ = 470 km s$^{-1}$ between the positional centroids of the 
metal absorption features and the Ly$\alpha$ emission line,
we apply an offset correction \citep{Adelberger:03} and measure a systemic
redshift of $z=4.866$ for the galaxy.
Interestingly, Ly$\alpha$ absorption immediately shortward of the emission line
is not obviously damped at our spectral resolution, 
with a column density of 
log [$N_{HI}$ (cm$^{-2})] < 20.0 \pm 0.3$ dex (see Figure 2 inset).  
At our spectral resolution, the detected absorption lines are strongly-saturated
and thus are not suitable for deriving metallicities.   
The most unusual feature of the spectrum is the broad absorption
trough in the Ly$\alpha$ forest first appearing at the Ly$\alpha$
emission feature and extending towards shorter wavelengths.  Although
two transmission spikes are visible at $\sim$6800 \AA \ and $\sim$7040 \AA,
nearly 100 \% of the continuum is absorbed between wavelengths of 6850 - 7100 \AA.


What might this absorption represent?  
It is likely that we are detecting the presence
of many more overlapping Ly$\alpha$ forest clouds closer to the galaxy than
can survive the UV radiation field of QSOs at comparable distances.
At the resolution of our data, individual Ly$\alpha$ forest lines cannot be
resolved, and we would see only the effects of a large number of blended lines.
It is also possible that the
enhanced absorption could be from only a few overlapping, but very
high-$N_{HI}$ clouds, but we might then expect to see more metal
absorption lines from these clouds redward of the Lya emission line,
which are absent in our data.
Even less likely is that this H~I overdensity signals the late completion of 
cosmic reionization, as recent observational studies have set convincing
constraints for this epoch at $z \apg$ 6.3 (Fan et al. 2006; Songaila 2004).   

We explore the nature of this remarkably large absorption trough
with a comparison
to the Ly$\alpha$ absorption seen towards their brighter counterpart objects, the QSOs.
QSOs produce radiation fields that
 ionize hydrogen over large, $\sim$16 physical Mpc regions at
 $z=5$ (Fan et al. 2006) (red solid line in Figure~2 inset).
Although evidence has emerged of a softening of this proximity effect in the transverse 
direction
on small scales of $\apll 1.5$ Mpc from studies of QSO pairs (Bowen et al. 20006; Tytler et al. 2007). 
detailed studies of the Lyman-series forest are made routinely only outside the
ionizing influence of the QSOs.  In contrast, ordinary star forming galaxies
have modest proximity zones of size $ 0.1 h^{-1}$ physical Mpc \citep{Adelberger:03}, 
thereby enabling the study of H I and its structure in the vicinity of deep potential wells
where the gas is not ionized.  Despite the advantage of a small proximity zone,
galaxies suffer from being much fainter in the ultraviolet.  
Unlike the vast majority of star-forming galaxies,
our target is bright because it is strongly-lensed, thereby yielding a
spectrum suitable for measuring the transmitted flux $T$ in the Ly$\alpha$ forest 
in the proximity of a {\it galaxy}.

\section{H I Flux Transmission}

We calculate the Gunn-Peterson (GP) optical depth \citep{Gunn:65}, 
$\tau_{GP}^{eff} = -ln (T)$, 
where $T$ is the ratio of the average observed flux
 to the average unabsorbed continuum flux, $T = <f_{\lambda}/f_{cont}>$,
 and $f_{cont}$ is determined by stellar synthesis models fit to the 
 photometric data.  We compute
$\tau_{GP}^{eff}$ in several redshift bins  extending  
from a wavelength clear of the red wing of Ly$\beta$ up to the blue edge of
our model fit to the H~I in the source (blue dashed line in Figure~2 inset delimits
the extent of the highest redshift bin). 
Uncertainties in $\tau_{GP}^{eff}$ are dominated
by the intrinsic scatter of the continuum flux levels due to the stochastic
nature of the absorption in the IGM, or sample variance \citep{Tepper-Garcia:08},
but also include 
continuum placement errors and shot noise.    
Figure~3 shows the results measured from the $A1689\_7.1$ 
spectrum and compares our values for $\tau_{GP}^{eff}$ to those
observed toward a large sample of QSOs (Fan et al. 2006; Songaila 2004).

\begin{figure}
\includegraphics[viewport=42 0 300 345, scale=0.54]{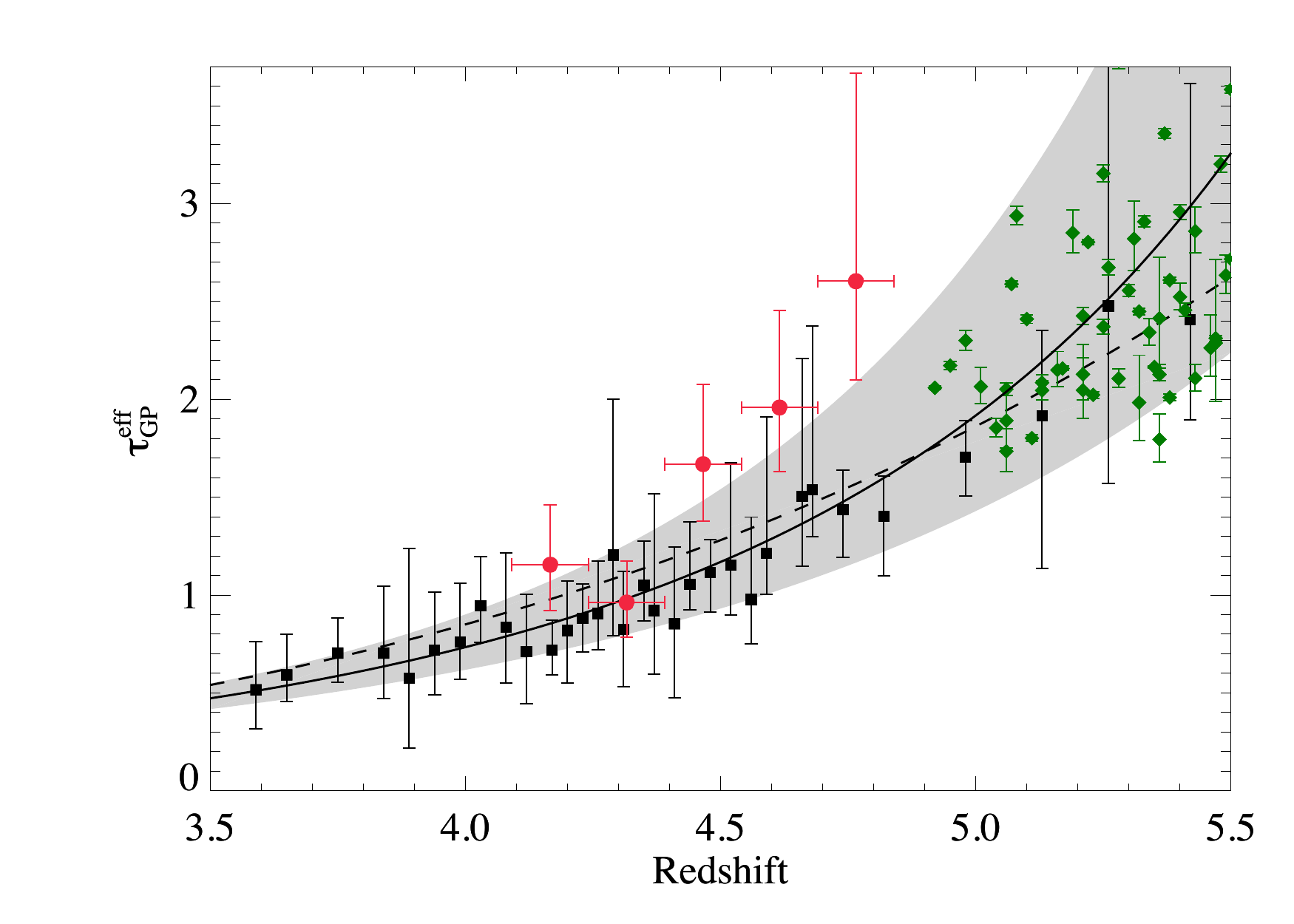}
\caption{H I overdensities in the high redshift universe.  The Gunn-Peterson effective 
optical depth, $\tau_{GP}^{eff}$, is measured in the Ly$\alpha$ forest of the spectrum for 
$A1689\_7.1$.  The values for $\tau_{GP}^{eff}$ in five redshift bins of $\Delta z=0.15$ 
each in the rest frame of the source (red points), are compared to measurements
towards QSOs [square-shaped points \citep{Songaila:04} and diamond-shaped 
points (Fan et al. 2006)].  While $\tau_{GP}^{eff}$ is expected to lie close to the 
bulk of the QSO points and model predictions [dashed curve (Fan et al. 2006) and 
solid curve \citep{Becker:07}],  surprisingly,  there is an excess in  $\tau_{GP}^{eff}$ 
along the line of sight to $A1689\_7.1$ at $z>4.5$.  The highest redshift point is the  
most deviant; it corresponds to a physical region of size 14 (physical) Mpc and this 
suggests that there is more H~I gas close to the galaxy compared to the standard IGM.
The error bars in the data are dominated by the intrinsic scatter of the continuum flux 
levels due to the stochastic nature of the absorption in the IGM, or sample variance, 
and also include continuum-placement uncertainties and shot noise. \label{fig_tau}}
\end{figure}

We emphasize that 
the values for $\tau^{eff}_{GP}$ are measured in the same way towards
both the QSOs and the galaxy, and after first excluding
the proximity zone, thereby yielding information only on the 
pervasive IGM; that is the IGM immediately {\it outside}
the photoionizing influence of the source.  Assuming $\tau^{eff}_{GP}$
to be the same in the standard IGM towards {\it any} background
object, we would expect $\tau^{eff}_{GP}$ to rise steadily towards
$A1689\_7.1$ with absorption redshift following the bulk of the QSO points. 
Compared to the predictions of a power-law model based on the density 
distribution (Fan et al. 2006) and a lognormal optical
depth distribution (Becker et al. 2007), $\tau_{GP}^{eff}$ measurements toward
$A1689\_7.1$ are in good agreement at $z < 4.5$; at $z>4.5$, however,
we see a significant excess in $\tau_{GP}^{eff}$.  If the underlying
mass distribution were unaltered by the presence of the galaxy, then this
highest-redshift point, the one in closest physical proximity of the galaxy, 
should have also followed the behavior of the other points.
The highest redshift point is the most deviant; it corresponds to 
a physical radius $\sim14$ (physical) Mpc, and this result
suggests that there is more H~I gas close to the galaxy compared to the
standard IGM.  

\section{Population Synthesis Modeling}
 $A1689\_7.1$ is imaged in
several bands, as follows: $i_{775} = 23.10 \pm 0.01$ (HST ACS), $J_{110}=23.10 \pm 0.02$ 
(HST NICMOS), $H = 23.45 \pm 0.38$ and $K_s = 23.45 \pm 0.35$
(Son of Isaac Instrument on ESO New Technology Telescope), and $P_{3.6}=23.3 \pm 0.02$,
and  $P_{4.5} = 23.34 \pm 0.03$ (IRAC on Spitzer Space Telecope).  
The observing details for these data are discussed elsewhere (Frye et al. 2007).
From these data we  set  initial constraints on the galaxy age and dust
extinction.
Our high quality $J_{110}$, $P_{3.6}$ and $P_{4.5}$ points
indicate only a modest 
H~I Balmer series continuum break at rest-frame $\sim$4000 \AA \, and thus a young
underlying stellar population dominated by O stars
with a minimum age of $t\apg10$ Myr (Figure 4).  The age of the
universe for our adopted cosmology sets the upper limit on the galaxy age of 1.2 Gyr.
  The dust extinction, as parameterized by the color excess $E(B-V)$, is fit with
 a broad range 
  spectral energy distribution
templates.   From these fits it is found 
that any model with $E(B-V)>0.1$ provides a poor fit to the slope of data,
as the continuum becomes significantly flatter in the blue with the addition of even moderate amounts of dust.  Thus we restrict our parameter space to $E(B-V) \leq 0.1$.

\begin{figure}
\includegraphics[viewport= 10 -10 150 280, scale=0.67]{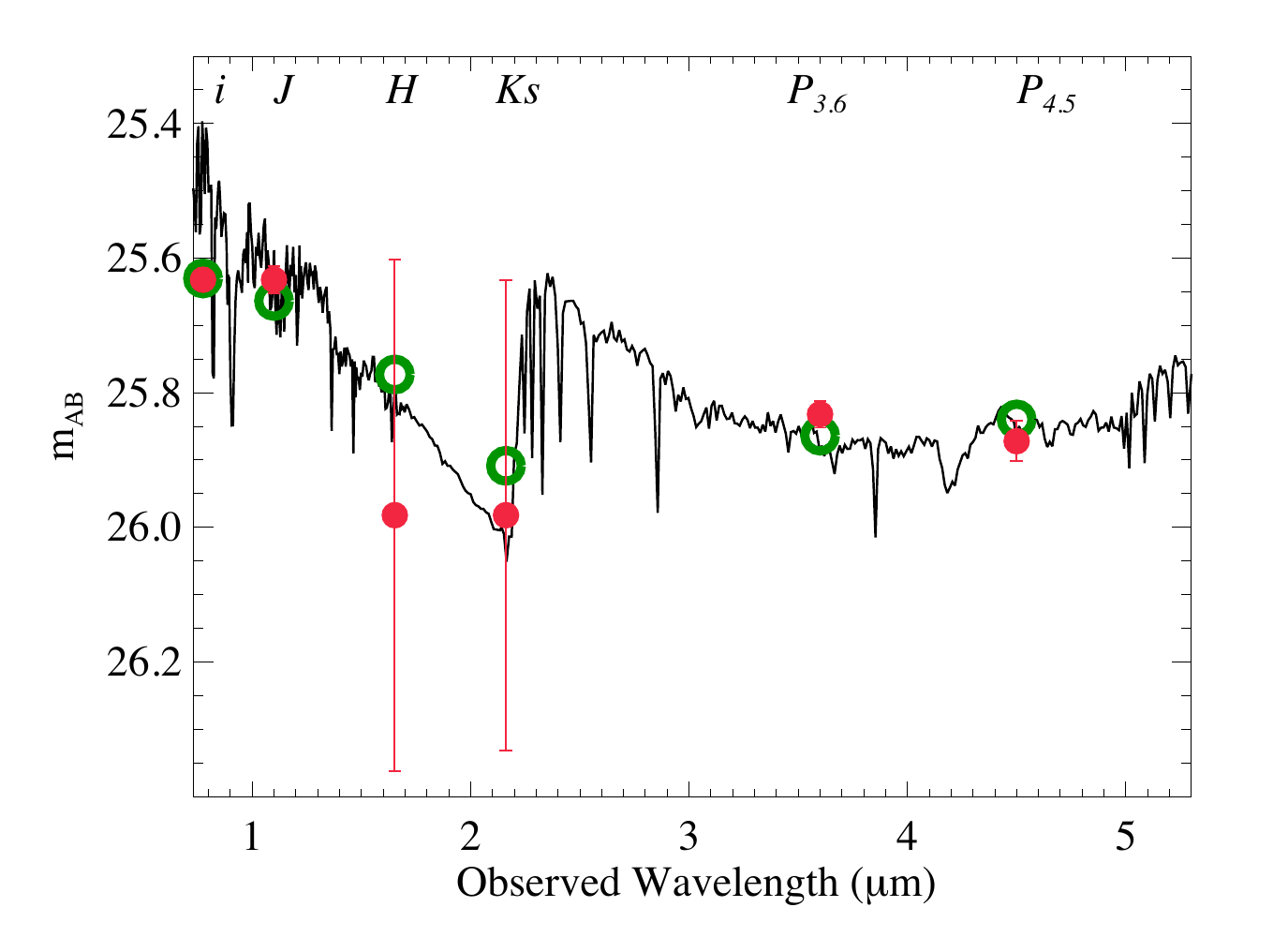}
\caption{
The best-fit spectral energy distribution for our galaxy, with dust extinction and attenuation
by the IGM.  The red solid circles indicate the values for our observed photometry
in the various filter bandpasses, as marked. 
We construct a suite of models and perform synthetic photometry on the models
(green open circles) until a model is found for which 
the reduced $\chi_r^2$ statistic is minimized with respect to the observed photometry.  
Our best-fit model is for a young, low-mass and dust-free galaxy.
\label{fig_bc}}
\end{figure}

Model spectral energy distributions are generated using the stellar synthesis
code of \citet{Bruzual:03}.
We select a Chabrier initial mass function (Padova 1994), stellar evolution tracks, and solar metallicity.
We choose a
 single starburst model with a range of decay rates $\tau$ and a star formation rate 
 (SFR) that depends exponentially on $\tau$ as follows:
$SFR(t) \propto exp(t/\tau)$ with $\tau =$ 0.1, 0.2, 0.3, 0.5, 1,
and 1.2 Gyr.  Continuous star-formation models are also considered, as approximated
by selecting $\tau$ to be
the age of the universe at $z=5$, and we do not consider more complicated star formation histories.  For each $\tau$, three parameters remain to be fit to the data: 
age $t$, $E(B-V)$, and stellar mass $M_*$.
A suite of models are constructed over the allowable parameter space for these
three variables, and each model is 
corrected for  the effects of dust extinction \citep{Calzetti:00},
and attenuation by the intervening Lyman-series forest \citep{Madau:95}.


We compute synthetic photometry on our models by
a  convolution of the model spectra with the observed filter bandpass transmission functions.
We compare the synthetic photometry with the observed photometric values 
until the lowest value of reduced $\chi^2$ is obtained.
Figure~\ref{fig_bc} shows our best fit model to the joint variation of  the parameters of
dust extinction, age, and mass, with values:  $E(B-V) = 0.0^{+0.02}_{-0.00}$, $t = 100_{-14}^{+14}$ Myr, and  $M^* = 7.3 \pm 0.70 \times 10^8 M_\odot$.  Note that as the exponentially-decaying models are normalized to have a total 
mass of 1$M_\odot$ as $t \to \infty$, 
we obtain the mass for $A1689\_7.1$ by the flux normalization that yields the lowest value for the reduced
 $\chi^2$ fit.
Our best fit is for a no-dust model.  
We compute formal $\chi^2$ uncertainties that represent 68 \% 
uncertainties for our input model.  We follow the prescription in
\citet{Cash:76} originally developed for use with X-ray spectral data that allows
for the joint estimation of confidence intervals.

The best fit  synthetic spectrum  is also shown in Figure~2 in the main text, but
{\it without} corrections for dust extinction and absorption by the intervening Ly$\alpha$ forest.  
This intrinsic form is used to define the unabsorbed continuum level $f_{cont}$ 
for $A1689\_7.1$ against which the flux transmission of the Ly$\alpha$
forest is measured.  

\section{Discussion}

Despite the strong magnification, the intrinsic luminosity
of the galaxy is not high; we measure an unabsorbed luminosity of
$L_{1400} = 7.7 \times 10^{28}$ ergs s$^{-1}$ Hz$^{-1}$, and an unlensed
$K = 24.5$, two magnitudes fainter than $K^*$ at $z=3$ \citep{Shapley:01}.
Galaxies at $z \apg 5$ are faint, and hence only a few high quality spectra
exist (Price et al. 2007; Kawai et al. 2006; Dow-Hygelund et al. 2007; Franx et al. 1997).
Some of these are at $z > 6$ where $\tau_{GP}^{eff}$ is
expected to be high (from the QSO measurements), but the spectrum of the
gamma-ray burst GRB060510B at $z = 4.94$ (Price et al. 2007).
shows some similar characteristics to the spectrum of
$A1689\_7.1$.   The IGM transmission measured
over a broad redshift bin of the former is $T = 0.18$, similar to the values
observed toward QSOs. However, this broad average smears out the
structure in $\tau_{GP}^{eff}$ as a function of redshift.  The
GRB060510B spectrum shows evidence of higher
than expected GP optical depth extending beyond the region affected by
the damped Ly$\alpha$ absorption at the redshift of the GRB host. It would be
interesting to reanalyze this GRB spectrum using smaller redshift
bins.    H~I opacities were also measured  inside the proximity zones of QSOs; although
significantly ionized, the proximity zones are found to have neutral hydrogen fractions that
exceed theoretical expectations.  This indicates that at least some QSOs are
also found in regions of gaseous overdensity with large sizes of
$15h^{-1}$Mpc \citep{Guimaraes:07}.  

Might $A1689\_7.1$ still be in the process of accreting 
much of its mass from its overdense surrounding cosmic structure?
Theoretical models predict that H~I gas is funneled into young
galaxies via large-scale filamentary structures \citep{Keres:05, Birnboim:03}.  
We have presented here observational evidence that H~I column densities 
are higher than expected near one high-redshift galaxy.  Based on 
the large physical size implied by the H~I excess, it is
unlikely that this gas will accrete onto a single galaxy.  
Alternatively, post-starburst galaxies are known to drive high-velocity 
outflows with velocities of $\apll 2000$ km s$^{-1}$ in low-ionization 
stages such as Mg~II \citep{Tremonti:07}.  Given sufficient time, the observed 
excess of H~I optical depth could be explained by such an outflow;
however $A1689\_7.1$ is too young by a significant factor given the best 
fit stellar age of 100$\pm 14$ Myr.
On the other hand, QSOs are known to drive outflows with 
terminal velocities in excess of $10^4$ km s$^{-1}$.  $A1689\_7.1$ does not
show obvious signatures of being an AGN,  but even if there was a 
low-metallicity outflow undetected in our spectrum, the velocities 
would still fall short by more than a factor of ten of that required to 
explain the H~I excess.  As more data become available to study the neutral H~I absorption
towards the several recently discovered strongly-lensed LBGs, it may be found to be typical
for high-redshift objects, both galaxies {\it and} QSOs, to be located in regions of space with neutral
hydrogen gas fractions significantly larger than that of the pervasive IGM.
 This re-evaluation of the H~I structure in the densest regions
of the universe at these epochs will provide the necessary
calibration for modeling the formation and evolution of galaxies in the early universe.

\acknowledgments

This work is based on data from the HST ACS instrument, which was developed
 under NASA contract NAS5-32865, and on the Very Large Telescope FORS2 instrument.
 B. L. F. is supported by Science Foundation Ireland Research Frontiers 
 Programme Grant
 PHY008.  D. V. B. is funded through NASA Long-Term Space Astrophysics grant 
 NNG05GE26G.  The work of D. A. C. was carried out at Jet Propulsion Laboratory, California Institute of Technology, under a contract with NASA.  
 We thank Holland Ford, Garth Illingworth, Avi Loeb, Rogier Windhorst, and Sergey Cherkis for discussions,
 and Rychard Bouwens
 for the HST NICMOS $J$-band photometry.


{\it Facilities:} \facility{HST(ACS)}, \facility{VLT}.

\end{document}